\documentclass[preprint,showpacs,preprintnumbers,amsmath,amssymb,superscriptaddress]{revtex4}

\usepackage{epsf}
\usepackage{graphicx}
\usepackage{sidecap}

\usepackage{color}

\def \beq {\begin{equation}}
\def \eeq {\end{equation}}
\begin{document}

\title{{Fermi Surface Topology and Hot Spot Distribution in Kondo Lattice System CeB$_6$}}




\author{Madhab~Neupane}
\affiliation {Joseph Henry Laboratory and Department of Physics,
Princeton University, Princeton, New Jersey 08544, USA}
\affiliation {Condensed Matter and Magnet Science Group, Los Alamos National Laboratory, Los Alamos, NM 87545, USA}

\author{Nasser~Alidoust}\affiliation {Joseph Henry Laboratory and Department of Physics, Princeton University, Princeton, New Jersey 08544, USA}

\author{Ilya~Belopolski}
\affiliation {Joseph Henry Laboratory and Department of Physics, Princeton University, Princeton, New Jersey 08544, USA}

\author{Guang~Bian}\affiliation {Joseph Henry Laboratory and Department of Physics, Princeton University, Princeton, New Jersey 08544, USA}

\author{Su-Yang~Xu}
\affiliation {Joseph Henry Laboratory and Department of Physics, Princeton University, Princeton, New Jersey 08544, USA}

\author{Dae-Jeong Kim}
\affiliation {Department of Physics and Astronomy, University of California at Irvine, Irvine, CA 92697, USA}

\author{Pavel P. Shibayev}
\affiliation {Joseph Henry Laboratory and Department of Physics, Princeton University, Princeton, New Jersey 08544, USA}

\author{Daniel S. Sanchez}
\affiliation {Joseph Henry Laboratory and Department of Physics, Princeton University, Princeton, New Jersey 08544, USA}

\author{Hao Zheng}
\affiliation {Joseph Henry Laboratory and Department of Physics, Princeton University, Princeton, New Jersey 08544, USA}

\author{Tay-Rong Chang}
\affiliation{Department of Physics, National Tsing Hua University, Hsinchu 30013, Taiwan}

\author{Horng-Tay Jeng}
\affiliation{Department of Physics, National Tsing Hua University, Hsinchu 30013, Taiwan}
\affiliation{Institute of Physics, Academia Sinica, Taipei 11529, Taiwan}


\author{Peter S.~Riseborough}
\affiliation {Department of Physics, Temple University, Philadelphia, Pennsylvania 19122, USA}

\author{Hsin~Lin}
\affiliation {Graphene Research Centre, Department of Physics, National University of Singapore, Singapore 117542, Singapore}

\author{Arun~Bansil}
\affiliation {Department of Physics, Northeastern University,
Boston, Massachusetts 02115, USA}

\author{Tomasz~Durakiewicz}
\affiliation {Condensed Matter and Magnet Science Group, Los Alamos National Laboratory, Los Alamos, NM 87545, USA}

\author{Zachary  Fisk}
\affiliation {Department of Physics and Astronomy, University of California at Irvine, Irvine, CA 92697, USA}

\author{M.~Zahid~Hasan}
\affiliation {Joseph Henry Laboratory and Department of Physics,
Princeton University, Princeton, New Jersey 08544, USA}
\affiliation {Princeton Center for Complex Materials, Princeton University, Princeton, New Jersey 08544, USA}

\date{today}
\begin{abstract}

{Rare-earth hexaborides have attracted a considerable attention recently in connection to variety of correlated phenomena including heavy fermions, superconductivity and low temperature magnetic phases.
Here, we present high-resolution angle-resolved photoemission spectroscopy studies of trivalent CeB$_6$ and divalent BaB$_6$ rare-earth hexaborides. 
We find that the Fermi surface electronic structure of CeB$_6$ consists of large oval-shaped pockets around the $X$ points of the Brillouin zone, while the states around the zone center $\Gamma$ point are strongly renormalized. Our first-principles calculations agree with our experimental results around the $X$ points, but not around the $\Gamma$ point, indicating areas of strong renormalization located near $\Gamma$.
The Ce quasi-particle states participate in the formation of hotspots at the Fermi surface, while the incoherent $f$ states hybridize and lead to the emergence of dispersive features absent in the non-$f$ counterpart BaB$_6$. 
Our results provide a new understanding of the electronic structure in rare-earth hexaborides, which will be useful in elucidating  the nature of the exotic low-temperature phases in these materials.}


\end{abstract}
\date{\today}
\maketitle

Rare-earth hexaborides have attracted a considerable research interest recently in connection to variety of correlated phenomena including heavy fermions, superconductivity and hidden order phases \cite{Fisk, Coleman, Riseborough, Antonov}. Moreover, with the advent of topological insulators \cite{Hasan, SCZhang} an intense effort has started to search 
for symmetry-protected topological phases in correlated systems, where recently samarium hexaboride (SmB$_6$) has been predicted to be a topological Kondo insulator  \cite{Dzero, Dai}. 
Strong experimental evidence for a topologically non-trivial phase in SmB$_6$ has intensified these efforts even further.
Numerous photoemission and transport experiments have been performed to identify the existence of an odd number of in-gap surface states and a two-dimensional conductance channel at low temperatures in SmB$_6$ \cite{Neupane, DLFeng, Fisk_discovery, Fisk_1, Nan}. However, the surface states in the topological Kondo insulator (TKI) phase of SmB$_6$ only exist at very low temperatures \cite{Neupane} and their Fermi velocity is expected to be low due to a strong $f$-orbital contribution \cite{Dai, Neupane}, which limits their future applications in devices. Furthermore, a related rare-earth hexaboride YbB$_6$ has recently been considered as a novel correlated topological insulator without a Kondo mechanism \cite{YbB6_Neupane, YbB6_DLFeng}.  This correlated topological phase in YbB$_6$ has been proposed to be explainable by an adjustable correlation parameter (Hubbard-$U$) and a band inversion between the $d$ and $p$ bands under a nonzero Coulomb interaction value \cite{YbB6_Neupane}. Moreover, the rare-earth hexaborides can provide a platform to realize a rich variety of distinct electronic ground states such as ferromagnetic order in EuB$_6$ and superconductivity in LaB$_6$ ($T_c\sim0.5$ K) \cite{Dai_1, Antonov, resis}. 
  
Another member of the hexaboride family CeB$_6$ exhibiting non-superconducting heavy-fermion metallic behavior, has been intensely investigated in the past because of its intriguing low-temperature magnetic phases \cite {dhv, dhv_1, dhv_2, CeB6_mag, CeB6_mag_1, CeB6_mag_2, CeB6_mag_3, CeB6_neutron, CeB6_neutron_1} as well as dense Kondo behavior \cite{dense_Kondo_1, dense_Kondo_2, dense_Kondo_3}. CeB$_6$ exhibits
antiferromagnetic (AFM) order below T$_N$ = 2.3 K \cite{CeB6_neutron_1}, which is preceded by another phase transition at T$_Q$ = 3.2 K, whose order parameter has long remained hidden from standard experimental probes such as neutron diffraction \cite{CeB6_mag, CeB6_mag_1}. Recently, evidence for characteristic wave vectors Q$_{AFM}$ associated with an antiferromagnetic ordering in CeB$_6$ was provided in inelastic neutron scattering experiments \cite{CeB6_mag_3}. This study revealed that above the antiferromagnetic quadrupole ordering transition associated with a wave vector, there was more evidence for ferromagnetic interactions than for antiferromagnetic ones.
Despite these interesting aspects, basic experimental studies on the electronic groundstates of CeB$_6$ are almost entirely lacking.

\begin{figure}
\centering
\includegraphics[width=13.50cm]{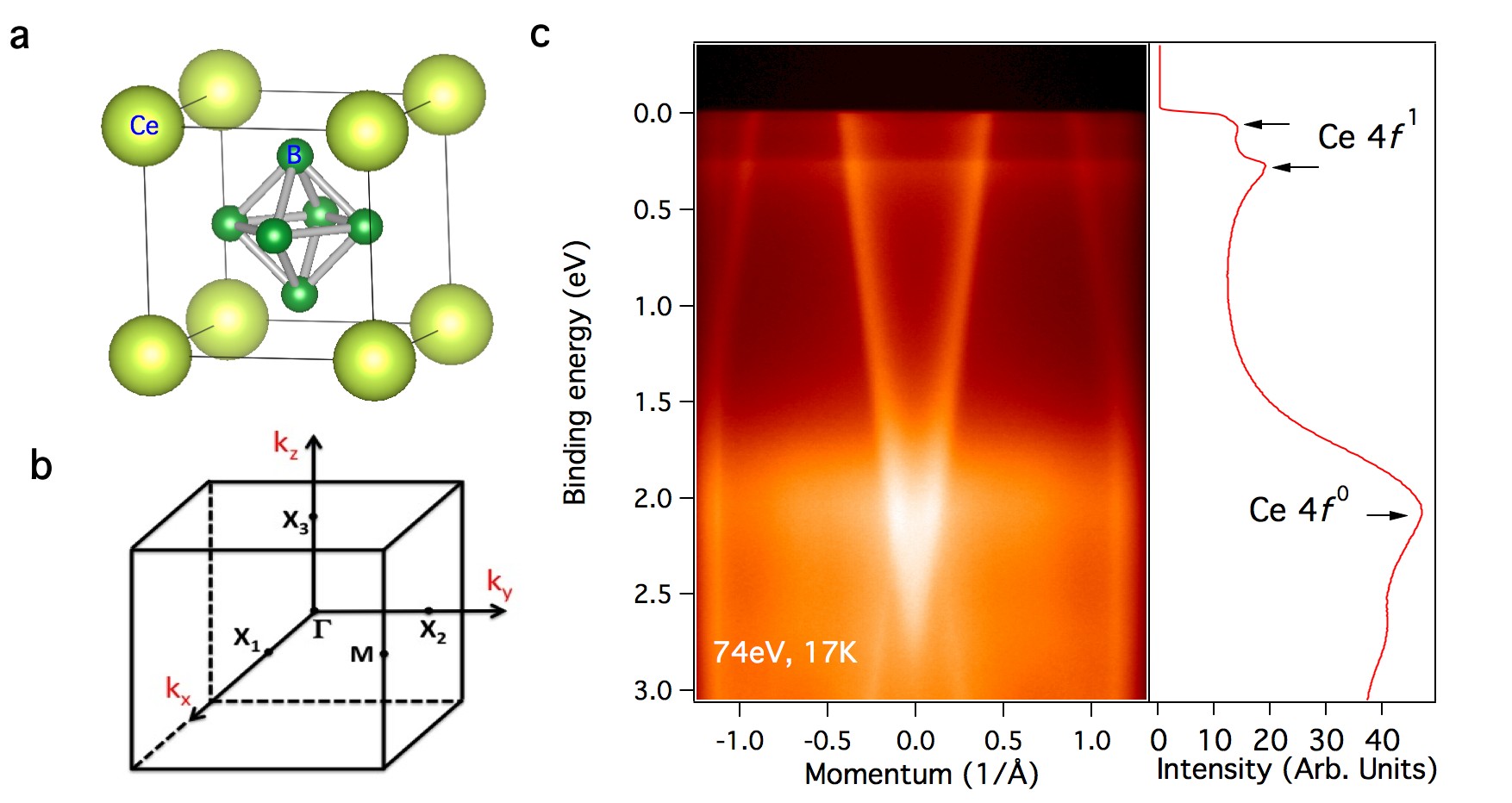}
\caption{{Crystal structure and electronic band structure of CeB$_6$}. (a) Crystal structure of CeB$_6$. Ce ions and B$_6$ octahedra are located at the corners and the center of the cubic lattice structure. \textbf(b) The bulk Brillouin zone of CeB$_6$. High-symmetry points are marked. (c) Band structure measured with ARPES along the ${M}$-${X}$-${M}$ momentum-space cut-direction (left) and the corresponding integrated energy distribution curve (right). Dispersive Ce 5$d$ band and non-dispersive flat Ce 4$f^1$ as well as a broad Ce 4$f^0$ peak are observed. This spectrum is measured with photon energy of 74 eV at a temperature of 17 K at ALS beamline 4.0.3.} 
\end{figure}

Here we study the electronic structure of the heavy-fermion hexaboride metal CeB$_6$, which has crystal structure identical to SmB$_6$  and YbB$_6$. Our angle-resolved photoemission spectroscopy (ARPES) data reveal  the presence of 4$f$ flat bands and dispersive 5$d$ bands in the vicinity of the Fermi level in CeB$_6$. 
We find that the Fermi surface electronic structure of CeB$_6$ consists of large oval-shape pockets around the $X$ points of the Brillouin zone (BZ).
Furthermore, our observations show that the area around the $\Gamma$ point is strongly renormalized, as indicated by highly increased density of states associated with band renormalization, called ``hotspots" due to their vicinity to the Fermi surface \cite{tomasz_1, tomasz_2}. 
Comparison with our first-principles bulk band calculations indicate that the $\Gamma$ point is strongly renormalized, while the $X$ point area does not undergo strong renormalization. 
The absence of such momentum-dependent hotspots in BaB$_6$ shows that this phenomenon is not a generic feature of hexaborides but is related to the strong electron-electron correlations and magnetic order in CeB$_6$. 
These experimental and theoretical results provide a new understanding of rare-earth hexaboride materials.

Single crystal samples of CeB$_6$ and BaB$_6$ used in our measurements were grown in the Fisk lab at the University of California (Irvine) by the Al-flux method, which is detailed elsewhere \cite{Fisk_discovery,resis}. Synchrotron-based ARPES measurements of the electronic structure were performed at the Beamlines 4.0.3 and 10.0.1 of the Advanced Light Source, Berkeley, CA equipped with high-efficiency Scienta R8000 and R4000 electron analyzers and beamline I4 of MAX-LAB III, Lund, Sweden equipped with a SPECS Phoibos 100 analyzer. The energy resolution was 10-30 meV and the angular resolution was better than 0.2$^{\circ}$ for all synchrotron measurements. The samples were cleaved along the (001) plane and were measured in ultrahigh vacuum better than 10$^{-10}$ Torr.
 The first-principles bulk band calculations were performed based on the generalized gradient approximation (GGA) \cite{GGA} using the projector augmented-wave method \cite{PAW, PAW_1} as implemented in the VASP package \cite{VASP, VASP_1}. The experimental crystallographic structure was used \cite{expt} for the calculations. The spin-orbit coupling was included self-consistently in the electronic structure calculations with a 12$\times$12$\times$12 Monkhorst-Pack $k$-mesh.

\begin{figure*}
\centering
\includegraphics[width=18.0cm]{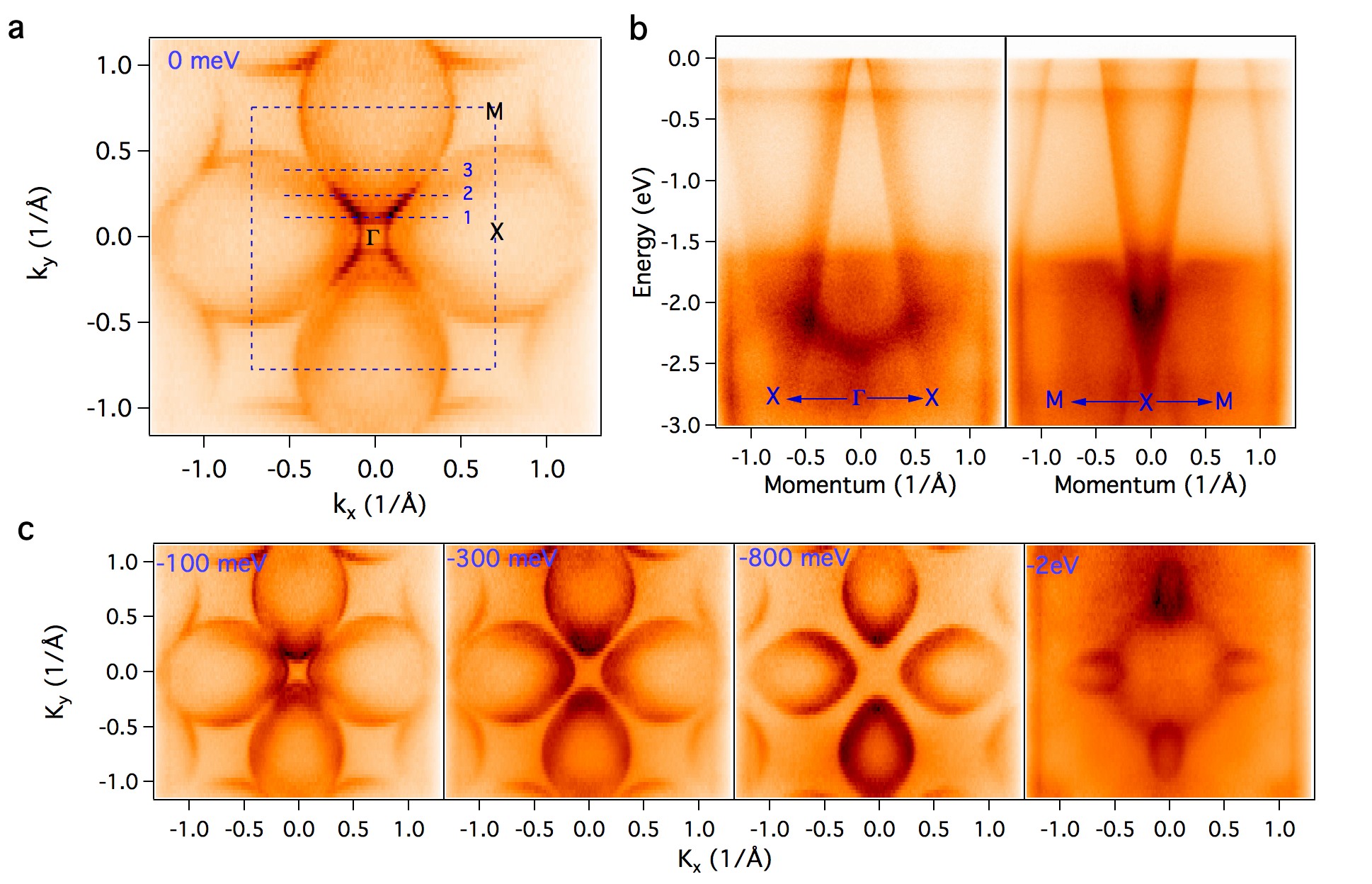}
\caption{{Distribution of hotspots in the electronic structure of CeB$_6$.}
(a) ARPES measured Fermi surface of CeB$_6$. Oval shaped pockets are observed at the ${X}$ points. Horizontal blue dashed lines and numbers correspond to the dispersion maps presented in Fig. 3a. (b) Band dispersion measured with ARPES along high-symmetry directions, which are marked on the plots. (c) ARPES measured constant energy contours. These data were collected at ALS BL 4.0.3 with a photon energy of 76 eV at a temperature of 17 K.}
\end{figure*}

\begin{figure}
\centering
\includegraphics[width=16.00cm]{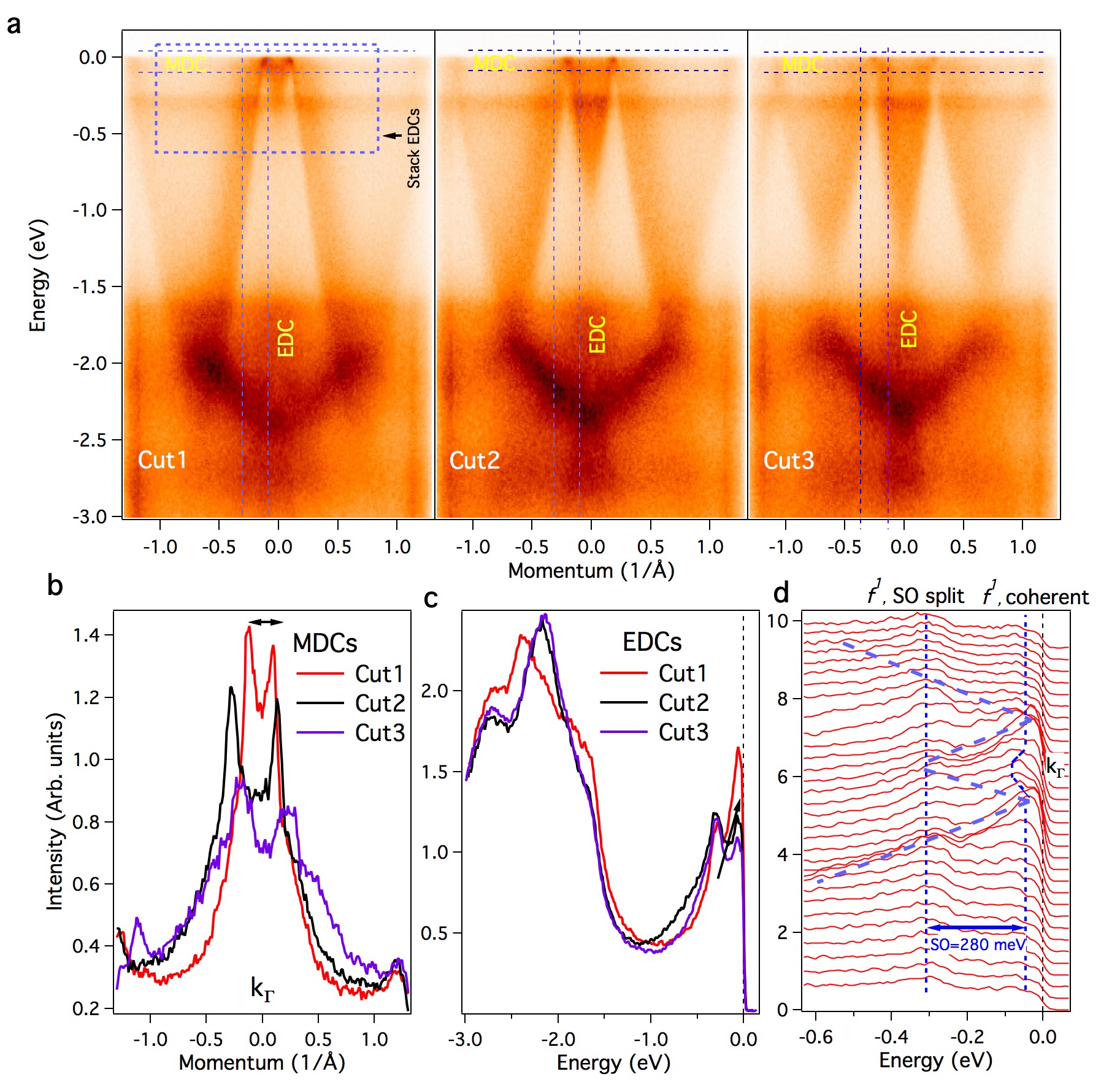}
\caption{{Evolution of hotspots in electronic structure.} (a) ARPES dispersion maps along Cut1, Cut2 and Cut3 as marked in Fig. 2(a). The vertical and horizontal parallel dashed lines represent the area of integration for MDCs and EDCs, respectively. The blue dashed rectangle denotes the area for stack EDCs. (b) MDCs along the cut directions marked in (a) by horizontal blue dashed parallel lines. (c) EDCs along the cut directions marked in (a) by vertical blue dashed parallel lines. The arrow in the vicinity of the Fermi level shows the shift of the spectral weight. (d) EDCs in the area marked by blue dashed rectangle in (a) for Cut1.}
\end{figure}

CeB$_6$ shares the same CsCl type crystal structure as SmB$_6$ and YbB$_6$, with the Ce ions and the B$_6$ octahedra being located at the corners and at the body center of the cubic lattice, respectively (Fig. 1a). 
CeB$_6$ is isostructural with BaB$_6$.
The bulk Brillouin zone (BZ) is cubic, where the center of the BZ is the $\Gamma$ point and the center of each face is the $X$ point (see Fig. 1b). In order to reveal the electronic properties of CeB$_6$, we systematically study its electronic structure at the (001) natural cleavage surface. To precisely determine the energy positions of the Ce $4f$ with respect to the Fermi level, we present a $k$-resolved dispersion map (Fig. 1c (left)). As it can be clearly seen in the integrated energy distribution curve of Fig. 1c (right), the two lowest $4f$ flat bands in CeB$_6$ are located near $E_F$ and approximately 0.3 eV below the Fermi level. level. These two features correspond to the spin-orbit split Ce 4$f$ of the $f^1$ configuration.
The broad peak located around binding energy of 2 eV also comes  from Ce 4$f$, but it represents the more localized $f^0$ configuration (see Fig. 1c (right)).




\begin{figure}
\centering
\includegraphics[width=15.0cm]{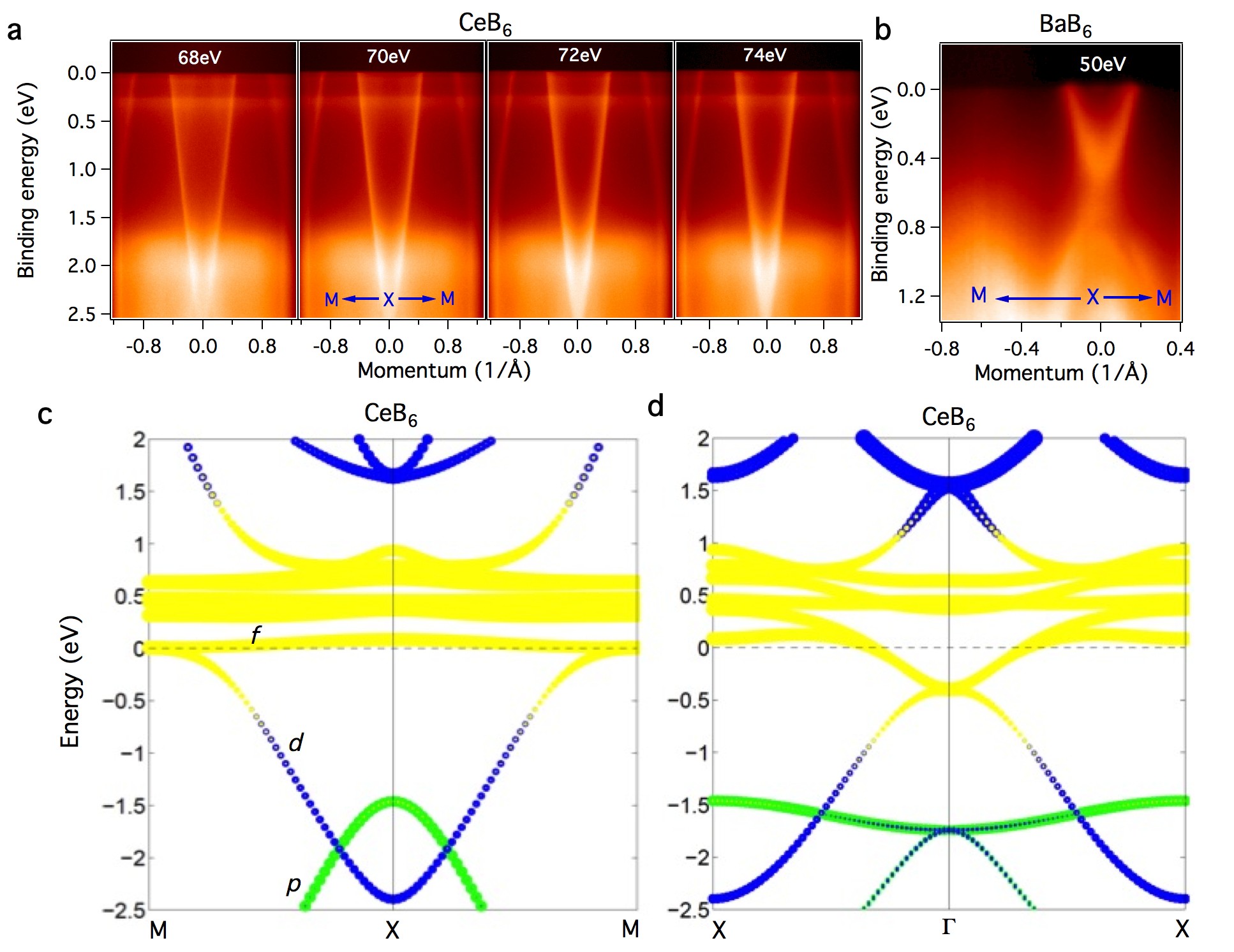}
\caption{{Photon energy dependent dispersion maps and first-principle electronic structure calculation of CeB$_6$.} (a) The photon energies  used in experiments are noted on the plots. The 4$f$ flat bands and dispersive 5$d$ bands are observed in the vicinity of the Fermi level.  These spectra are measured along the ${M}$-${{X}}$-${M}$ momentum space cut direction at a temperature of 17 K at ALS beamline 4.0.3. (b) ARPES dispersion map of BaB$_6$ along the high-symmetry direction, which is noted on the plot. The spectrum was measured at ALS beamline 10.0.1.
(c) Theoretical bulk band structure along the high-symmetry direction $M$-$X$-$M$, the maximum weight of $f$ bands are above the Fermi level, in contrast to the experimental data. (d) Calculated bulk bands along the $X$-$\Gamma$-$X$ high-symmetry direction.}
\end{figure}

In order to systematically resolve the low-energy electronic structure, in Figs. 2, 3 and 4a,b we present high-resolution ARPES measurements in the close vicinity of the Fermi level. 
Since the valence band maximum and conduction band minimum are located at the $X$ points, at the (001) surface one would expect the low energy electronic states to be located near the ${\Gamma}$ point and ${X}$ points.
The Fermi surface map of CeB$_6$ is presented in Fig. 2a, and reveals multiple pockets, which consist of an oval-shaped contour enclosing each ${X}$ point. The spectral intensity around the ${\Gamma}$-point is stronger as compared to that of the $X$ point. CeB$_6$ behaves as a Kondo metal in transport studies, which is consistent with our data. The low-energy electronic structure is mainly governed by the dispersive 5$d$ bands and flat 4$f$ bands as shown in the $k-$E maps along the high symmetry directions $X$-$\Gamma$-$X$ and $M$-$X$-$M$ of Fig. 2b. Here we also observe the bottom of the dispersive $d$-band to be about 2.5 eV below the Fermi level, and the dispersive $2p$ bands of Boron to be located in the vicinity of the bottom of this $d$-band.
In Fig. 2c, we present constant binding energy contours in the range from the vicinity of $E_{F}$ to $E_{B}$ = -2 eV. As can be seen in these figures the electron-like pockets around each of the ${X}$ points of the BZ grow in size upon going towards the Fermi level.
We note that de-Haas-van Alphen (dHvA) - derived Fermi surfaces are not revealing the structures around the $\Gamma$ point observed here with ARPES. It is not surprising, since as shown in \cite{dhv_1}, Fermi surface topology changes strongly with magnetic field, with the hotspot areas likely being the most affected, and hence dHvA produces similar shapes of Fermi surface for CeB$_6$ and e.g. LaB$_6$ \cite{dhv_3}. ARPES requires no magnetic field, and offers a novel view at these localized areas of enhanced density which might not be seen in dHvA.



In Fig. 3a, we present ARPES dispersion maps corresponding to the cut direction marked in Fig. 2a by blue dashed lines and integers.
Momentum distribution curves (MDCs) integrated within the horizontal parallel blue dashed lines in Fig. 3a are shown in Fig. 3b. 
The comparison of MDCs for three different cuts at different momenta shows the hotspot lines are localized in the area around the zone center encompassed by vector ${k}_\Gamma$, and their intensity is rapidly diminishing outside this region. Energy distribution curves (EDCs) integrated within the vertical parallel blue dashed lines in Fig. 3a are shown in Fig. 3c. 
Here, we note that the band renormalization driven by many-body effects, as shown above, leads to a maximum enhancement in the hotspot areas near the Fermi level, where bands in the vicinity, especially the top of the light band and the flat $f-$bands, provide extra phase space over which the interactions can lead to local accumulation of spectral weight. The interactions discussed here can be modeled in the future as a coupling of the electronic states to a bosonic mode associated with the magnetic fluctuations.
The shift in spectral weight, seen in the EDCs in Fig. 3d of the region marked by blue rectangle in Fig. 3a, is evidence for such a many-body mechanism. 
The dispersionless peak at 280 meV binding energy is the slightly hybridized spin-orbit split 
counterpart of the coherent part of the $f^1$ quasiparticle peak located near the Fermi level, which is indicated by blue dashed line close to $E_F$ in Fig. 3d. The coherent peak at the Fermi level is strongly hybridized with the light bands, and the renormalization, as observed by increased quasiparticle density and departure from generalized gradient approximation (GGA) calculation, is seen within the central area of Brillouin zone enclosed by vector ${k}_\Gamma$. Additional structures besides hotspots, like the small, shallow and relatively heavy electron-like pocket appear as a consequence of renormalization of the coherent $f^1$ band. 
Our data suggest that the interactions leading to a strongly momentum-dependent enhancement of the quasi-particle density are strong at the $\Gamma$ point but weak at the $X$ point.
This may be a characteristic sign of a fluctuation-driven band renormalization process, in which the coupling of the electronic and spin degrees of freedom can lead to a momentum-dependent Fermi surface renormalization \cite{Das}, which should be explored further in CeB$_6$. We note that such fluctuations are usually precursors to low-temperature phase transitions \cite{CeB6_neutron_1}, and they can occur at temperatures order of magnitude higher than transition temperature.


Our photon energy dependence measurements presented in Fig. 4a show the clear observation of 5$d$ bulk bands within the measured photon energy range of 68 eV to 74 eV (see Ref. \cite{SM} for wider range of photon energy data). The 4$f$ flat bands appear with no observable k$_{z}$ dispersion, as expected. These spectra were measured along the ${M}$-${{X}}$-${M}$ momentum space cut direction.
For comparison, we also measure the low-energy electronic structure of BaB$_6$ using ARPES. The dispersion map of this compound along the high-symmetry direction $M$-$X$-$M$ (Fig. 4b) shows the absence of any flat $f$ bands. The BaB$_6$ data is presented as a reference material \cite{SM}. Note that BaB$_6$ is measured in same experimental setup as CeB$_6$ and no hotspots are observed in the $d$-band of BaB$_6$, which indicates that hotspots are not generic features of $d$-band in the hexaborides, but rather arise from the important role played by electron-electron correlation and low temperature magnetic fluctuation. 
We note that magnetic fluctuations can typically persist up to temperatures order of magnitude larger than transition temperature \cite{magnetic_order}, which is also evidenced on the temperature dependent data \cite{SM}.


Finally, we present the first-principle electronic structure of CeB$_6$ along the high-symmetry directions $M$-$X$-$M$ and $X$-$\Gamma$-$X$ in Fig. 4c and 4d, respectively. These calculations which do not account for many-body interactions and resulting band renormalizations are in rough agreement with experimental results. 
Specifically, the experimental data show relatively good agreement with calculations in the vicinity of the $X$ point, but the agreement breaks down at the $\Gamma$ point, where we fail to see a hole-like pocket \cite{SM}. Instead, we find a strongly renormalized structure corresponding to hotspots on the Fermi surface. The origin of hotspots remains hidden, but we speculate that since magnetic fluctuations are precursors of magnetic instabilities, it is plausible that the fluctuations associated with the ferromagnetic instability \cite{CeB6_mag_3} build up already at the ARPES measurement temperature and it may be linked to the strongly momentum dependent hotspot structure, resulting from enhancement of the Bloch states in the vicinity of the $\Gamma$ point by the $Q=0$ magnetic fluctuations. 
The increased intensity at higher energies
is related to the onset of hybridization of conduction electrons with $f$-electrons. Moreover, the
strongly-momentum dependent parts are enhanced by coupling to fuctuations, which happen to have (1) a momentum dependence as also predicted from neutron scattering measurements \cite{CeB6_mag_3}
and (2) a temperature dependence ARPES data indicating rapid onset at low temperatures \cite{SM}.

In conclusion, by using high-resolution ARPES we have resolved the electronic band structure of CeB$_6$ which shows the presence of 4$f$ flat bands and dispersive 5$d$ bands in the vicinity of the Fermi level. We find that the Fermi level electronic structure consists of large oval-shaped pockets around the $X$ points of the BZ and highly normalize states surrounding the zone center $\Gamma$ point. 
We speculate that the hotspot observed in our data is linked to the unusual low-temperature order observed in this system. Specifically, the hotspot may be related to a high-temperature ferromagnetic fluctuation which is a precursor to magnetic order emerging at lower temperatures.
The absence of such a hotspot in BaB$_6$ shows that this phenomena is not a generic feature of hexaborides but is related to the strong electron-electron correlations and magnetic order in CeB$_6$. 
Our systematic experimental and theoretical results provide a new understanding of low temperature exotic phases of rare-earth hexaboride materials.




\bigskip
\bigskip
\textbf{Acknowledgements}
\newline
The work at Princeton is supported by the U.S. National Science Foundation Grant, NSF-DMR-1006492 and partial instrumentation support at Princeton is provided by the Gordon and Betty Moore Foundations EPiQS Initiative through Grant GBMF4547 (M.Z.H.). 
M.N. at LANL acknowledges the support by LANL LDRD program.
T.D. acknowledges support of the NSF IR/D program.
The work at Northeastern University is supported by the DOE, Office of Science, Basic Energy Sciences Grant Number DE-FG02-07ER46352, and benefited from Northeastern University's Advanced Scientific Computation Center (ASCC) and the NERSC supercomputing center through DOE Grant Number
DE-AC02-05CH11231. H.L. acknowledges the Singapore National Research Foundation for the support under NRF Award No. NRF-NRFF2013-03.
T.R.C. and H.T.J. are supported by the National Science Council, Taiwan. H.T.J. also thanks NCHC, CINC-NTU and NCTS, Taiwan, for technical support. P.R. was supported by the US Department of Energy, Office of Basic Energy Sciences through the award DEFG02-01ER45872.
We thank J. D. Denlinger for beamline assistance at the Advanced Light Source (ALS-LBNL) in Berkeley.
M.Z.H. acknowledges Visiting Scientist support from LBNL, Princeton University and the A. P. Sloan Foundation.
%
%




\setcounter{figure}{0}

\renewcommand{\figurename}{\textbf{Supplementary figure}}

\clearpage

\textbf{
\begin{center}
{\Large \underline{Supplementary Material}: \\
\vspace{0.15cm}
Fermi Surface Topology and Hot Spot Distribution in Kondo Lattice System CeB$_6$}
\end{center}
}

\vspace{0.2cm}

\begin{center}
Madhab Neupane, Nasser Alidoust, Ilya Belopolski, Guang Bian, Su-Yang Xu, Dae-Jeong Kim, Pavel P. Shibayev, Daniel S. Sanchez, Hao Zheng, Tay-Rong Chang, Horng-Tay Jeng,  Peter S.~Riseborough, Hsin Lin, Arun Bansil, Tomasz Durakiewicz, Zachary Fisk, and M. Zahid Hasan
\end{center}
\begin{center}
\textbf{This file includes:}
\end{center}

\vspace{0.15cm}

\textbf{
\begin{tabular}{l l}
\underline{I.} & Comparison of  calculations and experimental data\\
\underline{II.} & ARPES results on BaB$_6$ \\
\underline{III.} & Temperature and photon energy dependent spectra CeB$_6$\\
\end{tabular}
}
\vspace{0.35cm}

\textbf{Figs. S1 to S5}



\newpage

\textbf{I. Comparison of  calculations and experimental data}

We present a direct comparison between calculations and experimental data in Fig. S1. Fig. S1(a) shows the bulk band calculations along $M$-$X$-$M$ momentum space direction (left) and corresponding  experimental data (right). Similarly, Fig. S1(b) shows the calculations and ARPES data along the $X$-$\Gamma$-$X$ momentum space direction. These calculations which do not account for many-body interactions and resulting band renormalizations are in rough agreement with experimental results. 
More precisely, the experimental data shows relatively good agreement with calculations in the vicinity of the $X$ point, but the agreement breaks down at the $\Gamma$ point, where we fail to see a hole-like pocket. Instead, we observed a strongly renormalized structure corresponding to hotspots on the Fermi surface. 


\begin{figure*}
\includegraphics[width=16cm]{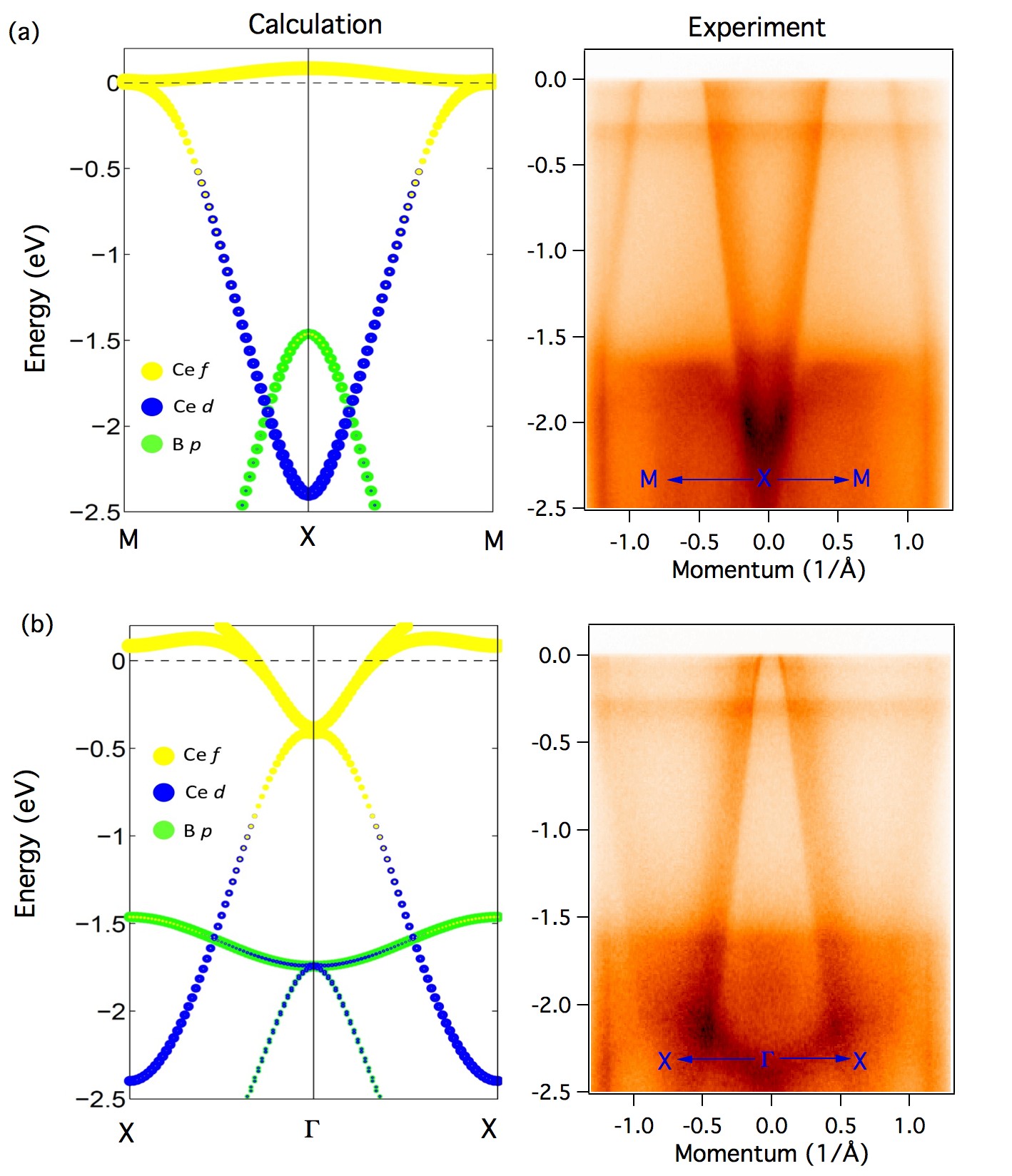}
\caption{{CeB$_6$: comparison of calculations and experimental data.
(a) Bulk band calculations along $M$-$X$-$M$ momentum space direction (left) and corresponding  experimental data (right). (b) Same as (a) along the $X$-$\Gamma$-$X$ momentum space direction. These data were measured at ALS BL 4.0.3 with photon energy of 76 eV and temperature of 17 K. }}
\end{figure*}


\vspace{0.5cm}

\textbf{II. ARPES results on BaB$_6$}

Figures S2 and S3 show the results of ARPES measurements: Fermi surface and band structure of BaB$_6$.  
Due to the plethora of $f$-electron specific spectral features that can be mistaken for effects
driven by non-$f$ orbitals, the practice of comparing $f$-bearing to non-$f$ materials
while taking ARPES data is of great use. Thus, the purpose of the BaB$_6$ measurement is to provide a material reference using another, non-$f$ member of the hexaboride family. 
Experimental result of Fermi surface map in Fig. S2  shows small circular pockets around the X points of the Brillouin zone. However, no states around the zone centre were observed. The BaB$_6$ data is presented as a reference material. The Fermi surface map of BaB$_6$ presented in Fig. S2 was measured in same experimental setup as CeB$_6$ and no hotspots are observed in the $d$-band of BaB$_6$.
Furthermore, we present a direct comparison of experimental data and theoretical calculations of BaB$_6$ in Fig. S3. Figure S3(a) shows the bulk band calculations along the $M$-$X$-$M$ momentum space direction (left) and corresponding  experimental data (right). Similarly, Fig. S3(b) shows the calculations and ARPES data along the $X$-$\Gamma$-$X$ momentum space direction.

\begin{figure*}
\includegraphics[width=16cm]{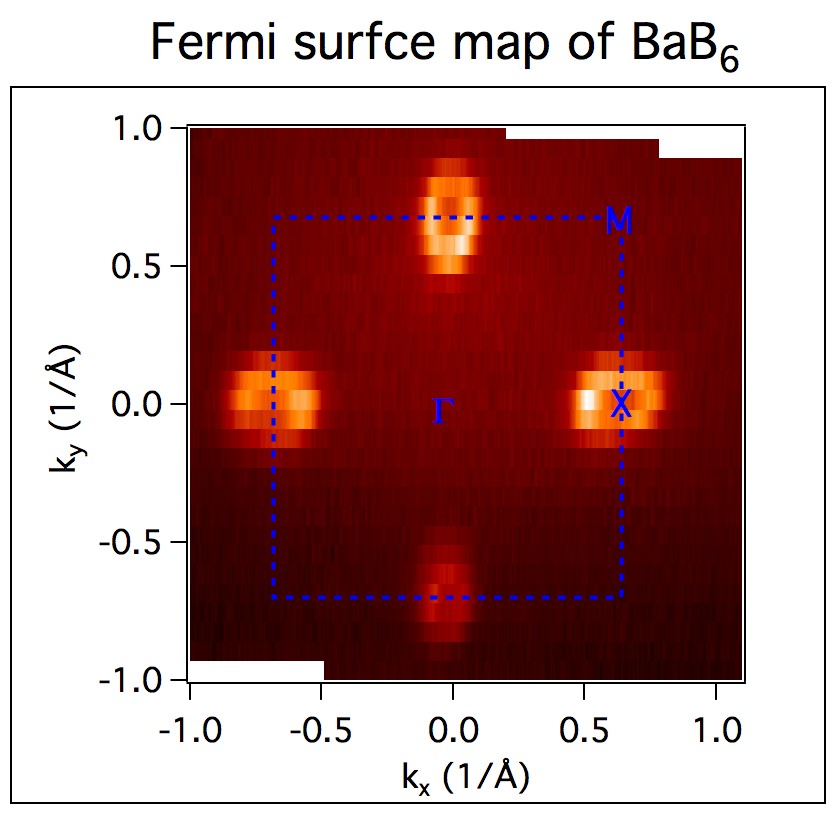}
\caption{{ARPES results on BaB$_6$.}  Fermi surface map of BaB$_6$ (left). These data were measured at ALS BL 4.0.3 with photon energy of 76 eV and temperature of 17 K.}
\end{figure*}

\begin{figure*}
\centering
\includegraphics[width=18cm]{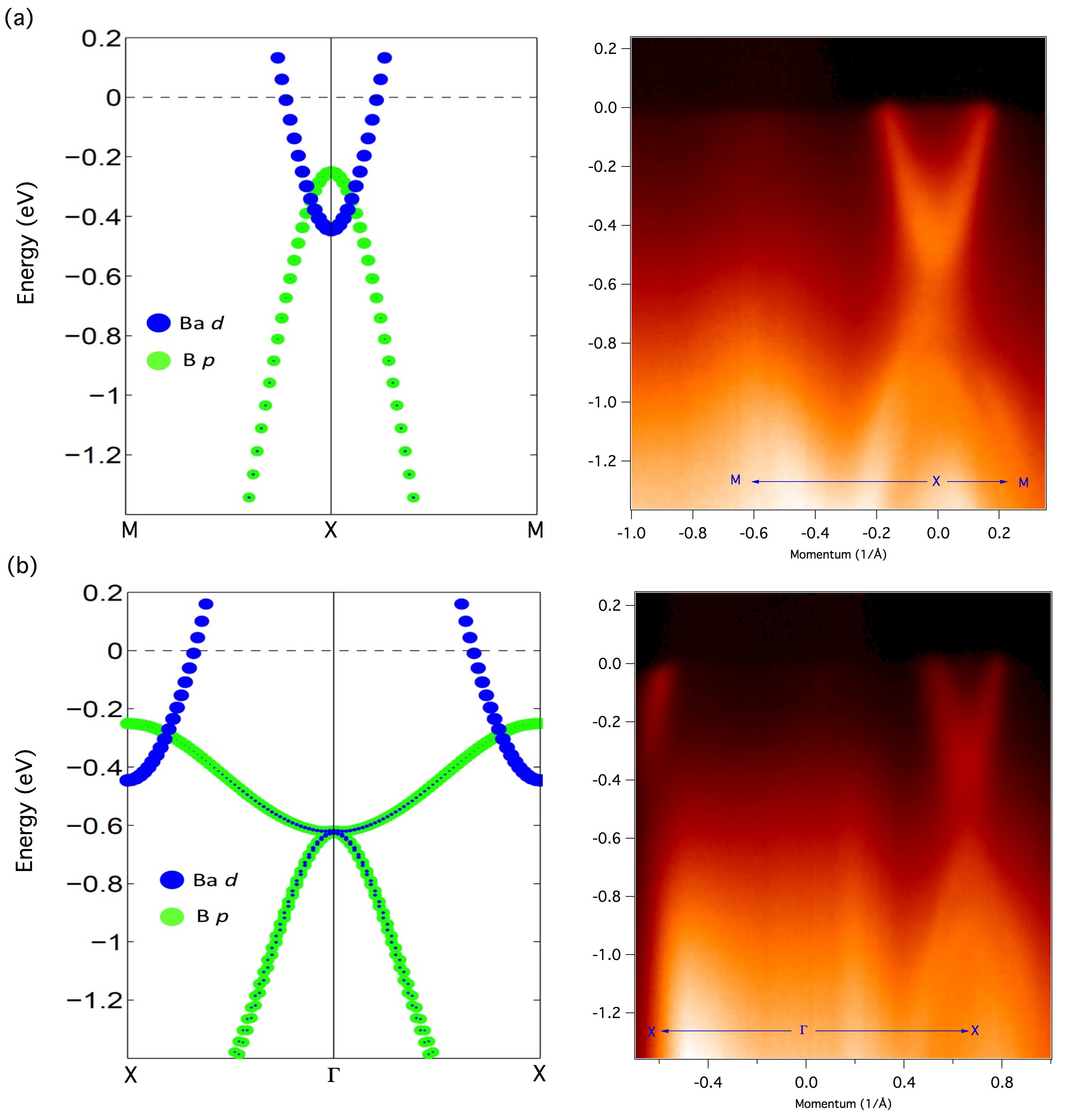}
\caption{{BaB$_6$: comparison of calculations and experimental data.
(a) Bulk band calculations along $M$-$X$-$M$ momentum space direction (left) and corresponding  experimental data (right). (b) Same as (a) along the $X$-$\Gamma$-$X$ momentum space direction. These data were measured at ALS BL 10.0.1 with photon energy of 50 eV and temperature of 15 K. }}
\end{figure*}


\vspace{0.5cm}


\textbf{III. Temperature and photon energy dependent spectra of  CeB$_6$}

\textbf{Temperature dependent photoemission spectra:} To check the temperature dependence of the electronic structure of the hotspots, we cleaved another CeB$_6$ sample at temperature of 30 K and measured temperature dependent Fermi surface maps at various temperatures. The Fermi surface map measured at 30 K is shown in Fig. S4 (left), where a much weaker hotspot spectral intensity can be observed as compared with 17 K results shown in Fig. 2 of main-text. These data were collected at 74 eV which is different than 76 eV data shown in the main-text.  The observation of a weak hotspot intensity here at 30 K with different photon energy rules out the possibility of the matrix element effect.  

Furthermore, by raising the temperature the hotspots spectral intensity near the zone centre is observed to be much weaker (see 60 K and 100 K Fermi surface maps in Fig. S4). 
Presence of slightly enhanced but temperature independent intensity at higher temperatures indicates the interactions present at low-temperatures, such as fluctuations at temperatures up to an order of magnitude above transition temperature, can be responsible for areas of enhanced quasiparticle weight.



\begin{figure*}
\includegraphics[width=16cm]{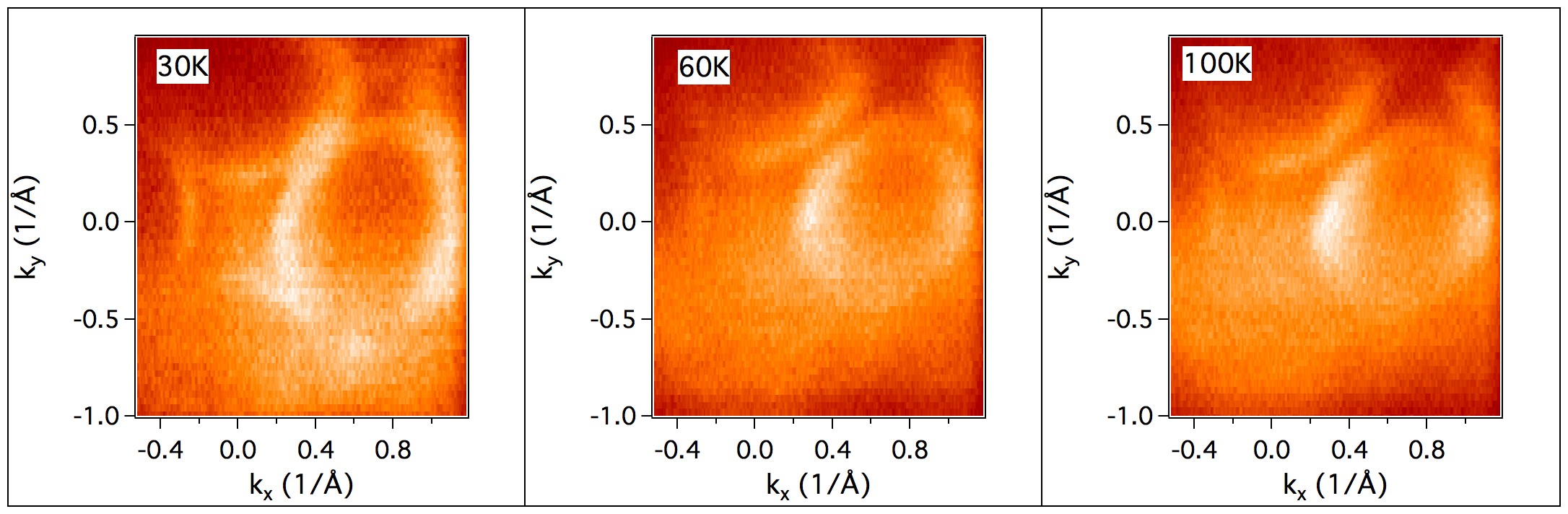}
\caption{{Temperature dependent ARPES results on CeB$_6$.}  Fermi surface map of CeB$_6$ measured at various temperatures. The measured temperatures are noted on the spectra. These data were measured at ALS BL 4.0.3 with photon energy of 74 eV.}
\end{figure*}

\begin{figure*}
\includegraphics[width=18cm]{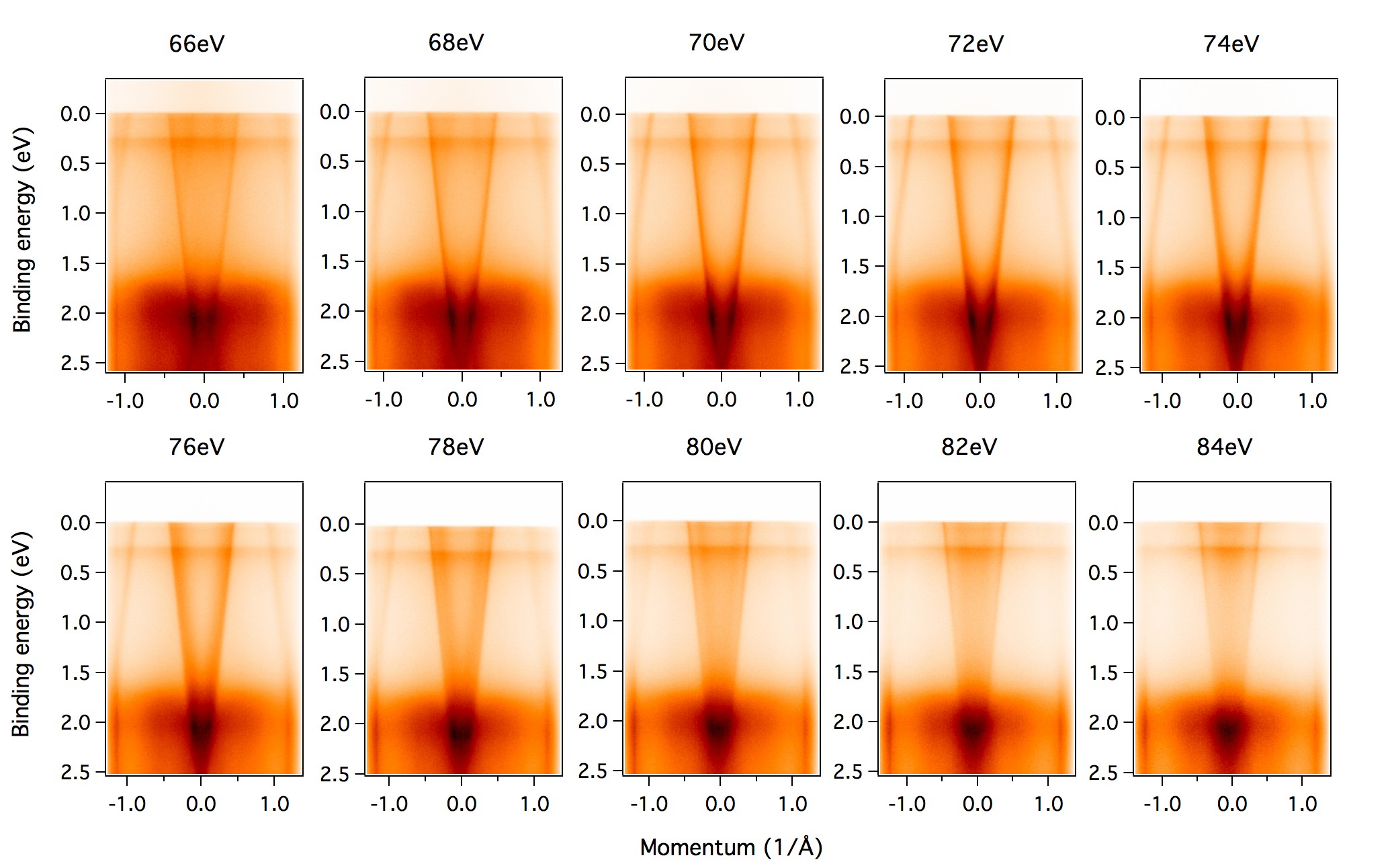}
\caption{{Photon energy dependent ARPES results of CeB$_6$.} Measured photon energies (from 66 eV to 84 eV) are marked on the plot. These data were measured at ALS BL 4.0.3 with temperature of 17 K.}
\end{figure*}

\textbf{Photon energy dependent photoemission spectra:} Figure S5 shows the dispersion maps measured using photon energy from 66 eV to 84 eV with 2 eV energy step. Using inner potential of 14 eV, our measurements cover a wider range of Brillouin zone from $k_z$ = 4.44 to 4.95 $\AA^{-1}$. We note that the $k_z$ = 4.5 corresponds close to the $\Gamma$ point.

\vspace{0.4cm}
Correspondence and requests for materials should be addressed to M.N. (mneupane@lanl.gov) and M.Z.H. (Email:
mzhasan@princeton.edu).

\end{document}